\documentclass[preprint,aps]{revtex4} 

\usepackage{amsmath}
\usepackage{graphicx}
\usepackage{gensymb}

\def\be{\begin{equation}}
\def\ee{\end{equation}}
\def \bea#1\eea {\begin{eqnarray}#1\end{eqnarray}}

\begin{document}

\title{Correspondence Between the Phase Diagrams of TIP5P Water \\ and a Spherically Symmetric Repulsive Ramp Potential
}

\author{Zhenyu Yan$^1$, Sergey V. Buldyrev$^{2,1}$, Pradeep Kumar$^1$, \\
Nicolas Giovambattista$^{3}$ and H. Eugene Stanley$^1$}

\affiliation{
$^1$Center for Polymer Studies and Department of Physics, Boston
University, Boston MA 02215, USA\\
$^2$Department of Physics, Yeshiva University, 500 West 185th Street,
New York, NY 10033 USA\\
$^3$Department of Chemical Engineering, Princeton University,
Princeton, New Jersey 08544-5263 USA
}

\date{\today}

\newpage

\begin{abstract}

We perform molecular dynamics simulations of a well-known water model (the TIP5P pair potential) and 
a simple liquid model (a two-scale repulsive ramp potential) 
to compare the regions of anomalous behavior in their phase diagrams. 
We select the parameters of the ramp potential by mapping it
to an effective pair potential derived from the TIP5P model.
We find that the regions of anomalous behavior in the phase diagrams
of both systems can be mapped onto each other if (i) pressure $P$ and temperature $T$
are replaced by $T-T_{C}$ and $P-P_{C}$,
respectively, where $(T_{C},P_{C})$ are the coordinates of the liquid-liquid critical point of the
corresponding system; and (ii) a single ramp particle corresponds
to two TIP5P molecules. We present heuristic arguments supporting point (ii). We also argue that
the water-like anomalies in the ramp potential
are due to the ability of the particles to reproduce, upon compression or heating,
the migration of water molecules from the second shell to its first shell.

\end{abstract}

%\pacs{PACS numbers: 64.60.Ak, 62.20.Dc, 46.25.-y}

\maketitle

Liquid water is peculiar as reflected by its thermodynamic and dynamic anomalies~\cite{pablogene,jeffrey01},
such as the density decrease upon isobaric cooling (density anomaly) and the increase
of diffusivity upon isothermal compression (diffusion anomaly).
It has been proposed that these anomalies may arise from a liquid-liquid critical point (LLCP) in the deeply
supercooled state of water~\cite{poole92}.
Several other liquids (e.g., silica, silicon, carbon, and phosphorous)
with local tetrahedral order~\cite{angellPCCP,shell02,molinero06} also may show water-like anomalies.
These anomalies of water and the LLCP can be
reproduced by simple liquids interacting via core-softened spherically symmetric potentials
which lack the strong orientational interaction expected in tetrahedral liquids
~\cite{Stell70,cho96,stillinger97,sadr98,Jagla99,predp05,wilding02,xu06,zyan,sharma06,olive06,errington06}.

\begin{figure}
\includegraphics[width=8cm]{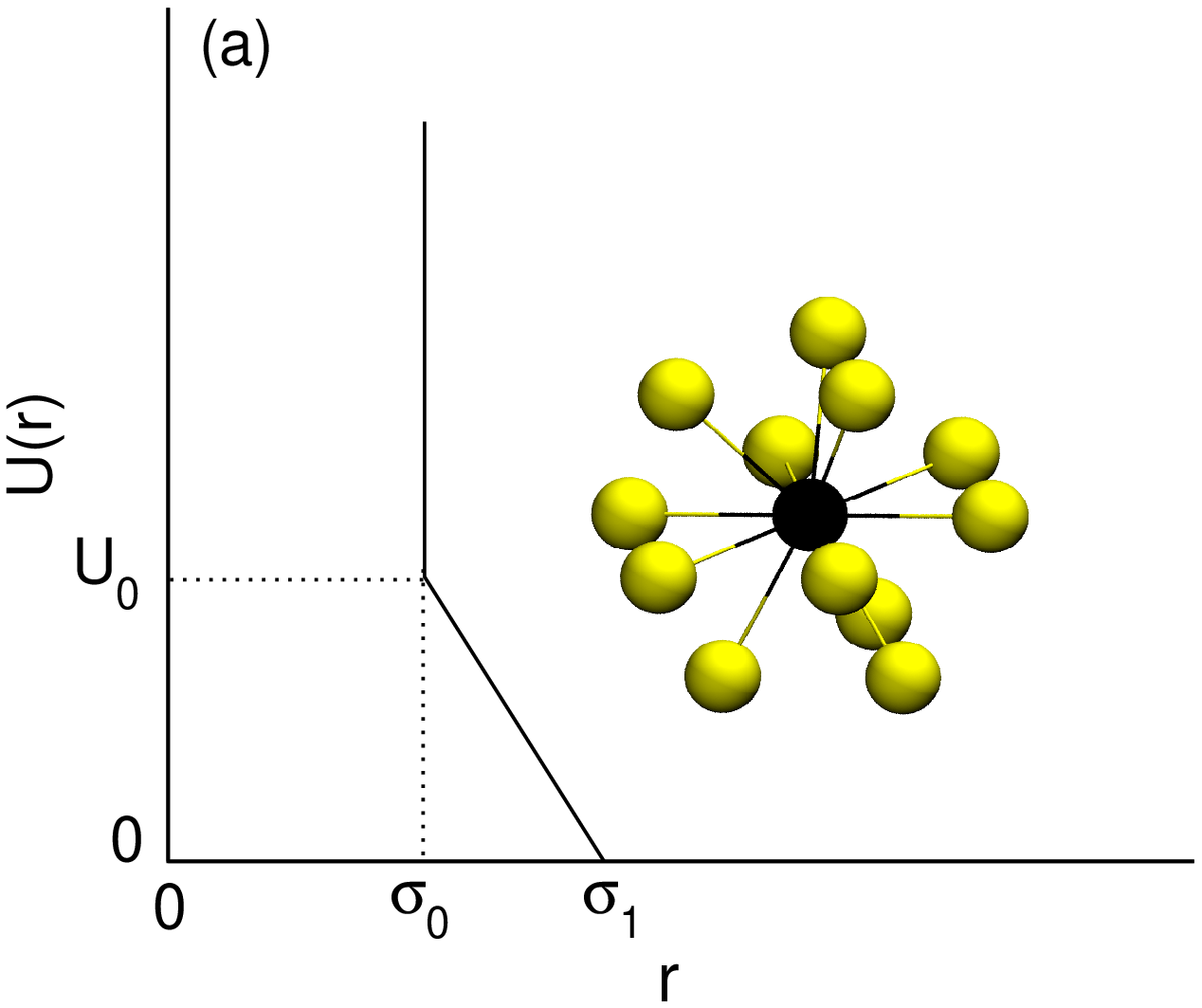} 
\includegraphics[width=8cm]{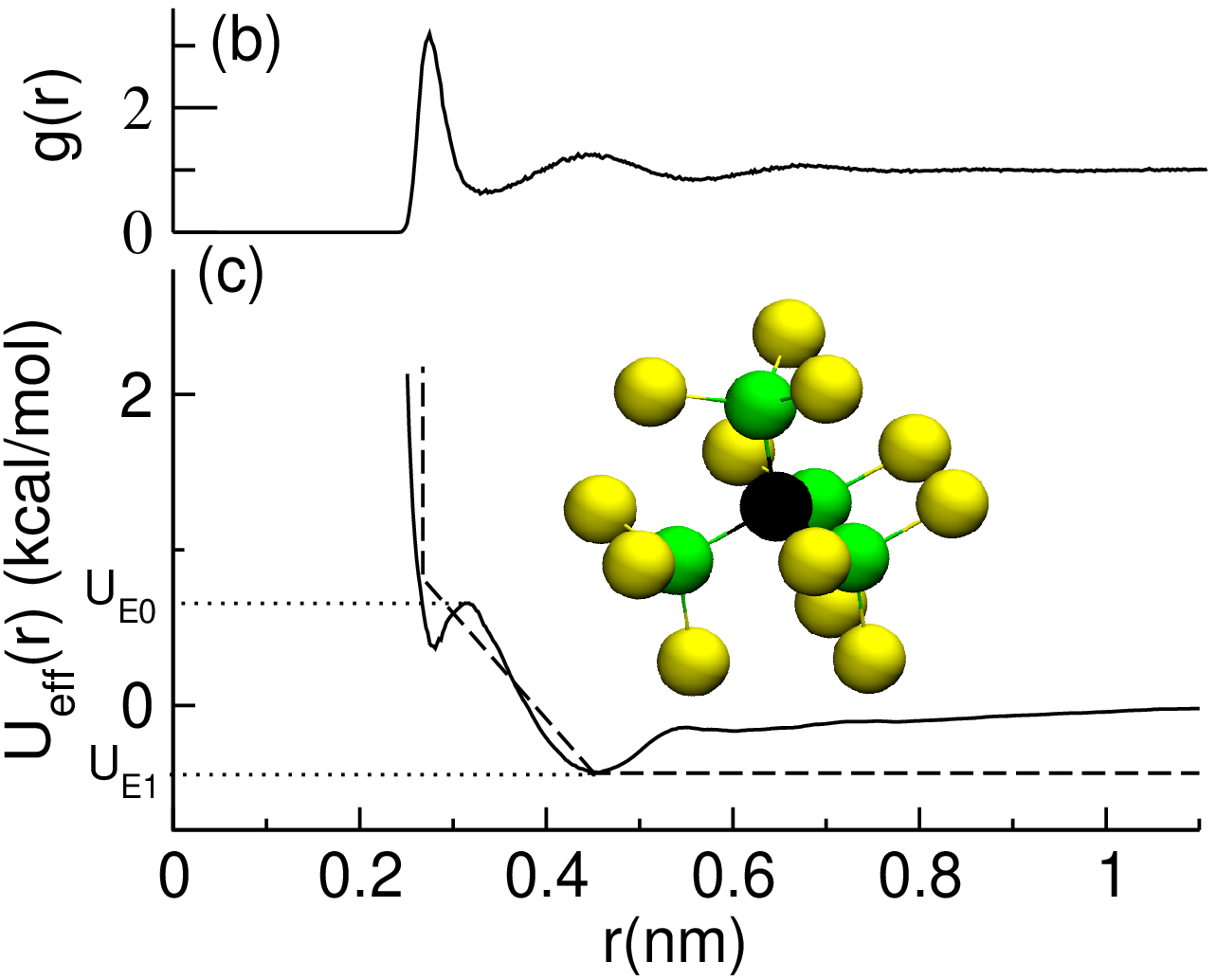}
\caption
{(Color online)
(a) The two-scale spherically symmetric repulsive ramp potential: $\sigma_0$ corresponds to the hard core distance,
$\sigma_1$ characterizes a softer repulsion range that can be overcome
at high $P$ and $T$. The central ramp particle (black) and its twelve nearest neighbors (yellow)
form a hcp crystal structure in a range of densities corresponding to the density anomaly.
In our study the parameters are defined in physical units as $\sigma_1=0.45$~nm, $\sigma_0=0.267$~nm and 
$U_0=1.31$~kcal/mol by mapping the ramp potential to the effective potential in (c) [see text].
(b) The pair correlation function, $g(r)$, and (c) spherically symmetric effective potential, $U_{\rm eff}(r)$, from the
simulations using the TIP5P model at $T=280K$ and $\rho=1.00$~g/cm$^3$ (solid line).
For hexagonal ice, the twelve neighbors (yellow) in the second shell of the center water molecule (black)
also has a hcp structure while the four nearest neighbors (green) in the first shell are located in the corner
of a tetrahedron.
$U_{\rm eff}(r)$ can be approximated by a ramp potential (dashed lines).
By calculating the area of $g(r)$ for $r \leq \sigma_0$ we find that the hard core of the ramp particle  
roughly incorporates two water molecules (see also Fig.~\ref{sergey}).
}
\label{pot}
\end{figure}

Water also possesses structural anomalies which occur when metrics describing
both translational and orientational order decrease upon compression,
as found in both the SPC/E and TIP5P water models~\cite{jeffrey01,zyan2007}. 
Water's structural anomaly is also reproduced by a family of
core-softened spherically symmetric potentials possessing two characteristic length scales
$\sigma_0$ and $\sigma_1$ (see the ramp potential in Fig.~\ref{pot}(a))~\cite{zyan}. 
In order to exhibit a water-like structural anomaly,
the ratio $\lambda\equiv\sigma_0/\sigma_1$
must lie within a small interval around 0.62, the ratio of the distances to water's first and
second neighbor shells, $0.28$~nm/$0.45$~nm\cite{zyan}. 

A quantitative connection between the ramp potential and water's pair
potential has not been established, as well as the relation between the
regions of anomalies in their respective phase diagrams. 
In this work, we show that the effective pair potential derived from
the TIP5P water model~\cite{tip5p} can be approximated by a two-scale 
spherically symmetric repulsive ramp potential, allowing us to
assign physical units to the temperature and pressure of the ramp model. 
We perform molecular dynamics simulations using both the TIP5P and ramp potentials and compare the 
regions of anomalies in the corresponding phase diagrams.
We find that the regions of anomalies in both phase diagrams are 
{\it quantitatively} similar if (i) pressure $P$ and temperature $T$ are measured in terms of $T-T_{C}$ and $P-P_{C}$,
respectively, where $(T_{C},P_{C})$ are the coordinates of the liquid-liquid critical point of the
corresponding system; and (ii) a ramp particle corresponds
to two TIP5P molecules. We present quantitative arguments supporting point (ii) and provide a simple
picture to explain the similarities observed in the TIP5P and ramp potentials. 
A ramp liquid particle corresponds {\it effectively} to two water molecules, one molecule plus $1/4$ of each of its
four neighbors. The water-like anomalies in the ramp potential
are due to the ability of the particles to reproduce, upon compression or heating, 
the migration of water molecules from the second shell to its first shell.

The TIP5P model is a well-known water model and its parameters are defined in physical units,
so values of $P$ and $T$ from simulations can be compared directly with experiments~\cite{tip5p}.
Instead, thermodynamic properties in the ramp potential are given in terms of potential parameters,
 such as $\{\sigma_0, U_0\}$, and the particle mass, $m$. 
To compare the phase diagrams of the ramp potential to that of the 
TIP5P model, we will define $\sigma_0$ and $U_0$ in units of `nm' and `kcal/mol', respectively, and
$m$ in units of `g/mol'. We do this by calculating $U_{\rm eff}(r)$,
the effective spherically symmetric pair potential between water molecules
from the TIP5P model simulations.
$U_{\rm eff}(r)$ is obtained from the oxygen-oxygen pair correlation
function $g(r)$, by solving the Ornstein-Zernike equation and using the
hypernetted chain approximation~\cite{stillinger93}.
The resulting $U_{\rm eff}(r)$ depends on $T$ and density $\rho$~\cite{johnson07}, but
has no significant change for different state points
in the anomalous region. For the TIP5P model, the range of anomalies is 
approximately $220$~K~$<T<320$~K and $0.90$~g/cm$^3$~$<\rho<1.16$~g/cm$^3$~\cite{zyan2007}.
We select a state point located in the middle of the anomalous regions, at $T=280$~K and $\rho=1.00$~g/cm$^3$,
and calculate $g(r)$ and $U_{\rm eff}(r)$ [see Fig.~\ref{pot}(b) and Fig.~\ref{pot}(c)].
We find that $U_{\rm eff}(r)$ is similar to the effective pair potential  
obtained from the experimental $g(r)$~\cite{stillinger93}, and
shows a hard-core-like steep repulsion at $r\approx0.26$~nm and 
an approximately linear repulsive region covering the distance spanned by the
second shell of a central water molecule, approximately $0.32$~nm~$ < r < 0.45$~nm.
The shallow minimum at $r=0.28$~nm is caused by hydrogen-bonding attraction
and corresponds to the first peak of $g(r)$, while the minimum at $r=0.45$~nm
[$U_{E1}\equiv U_{\rm eff}(0.45$~nm)$=-0.45$~kcal/mol]
corresponds to the second peak of $g(r)$. $U_{\rm eff}(r)$ also shows a maximum at 
$r \approx 0.32$~nm [$U_{E0}\equiv U_{\rm eff}(0.32$~nm)$=0.66$~kcal/mol]
that corresponds to the first minimum of $g(r)$.

Figure~\ref{pot}(c) also shows that a ramp potential is a good approximation to $U_{\rm eff}(r)$. 
In the figure we set $\sigma_1=0.45$~nm and define the ramp part of the potential such that it intersects the 
plot of $U_{\rm eff}(r)$ at $(U_{E0} + U_{E1})/2$.
The intersection of the ramp part of the potential with the hard core
of $U_{\rm eff}(r)$ is used to define $U_0$ and $\sigma_0$. This results in $\sigma_0= 0.267$~nm, which is
located between 0.28~nm, the first peak position of $g(r)$ and 0.26~nm, roughly the infinite repulsion part of $U_{\rm eff}(r)$.
Therefore, $\lambda \equiv \sigma_0/\sigma_1=0.593$ and $U_0 = U_{\rm eff}(\sigma_0) - U_{\rm eff}(\sigma_1)=1.31$~kcal/mol.
$U_0$ is approximately the energy barrier that water molecules need to overcome
to migrate from the second shell to the first shell positions in terms of the effective potential.
It is also roughly the energy that ramp particles need to overcome to reach the hard core distance.

\begin{figure}
\includegraphics[width=8cm]{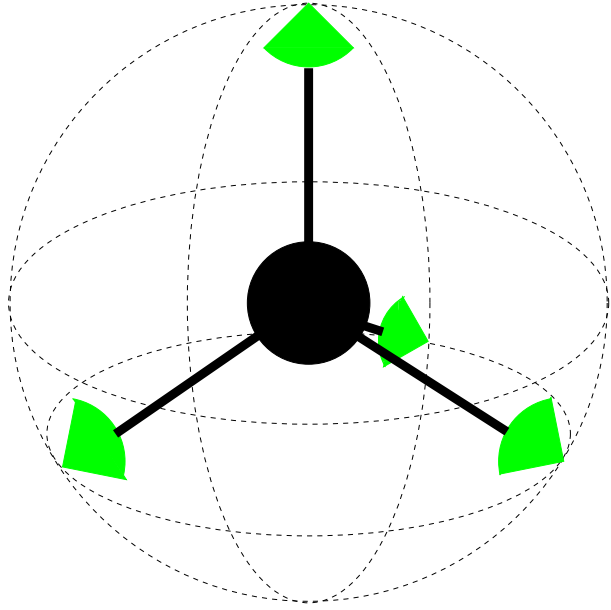}
\caption
{Sketch of a water molecule and $1/4$ of each of its four nearest neighbors in a tetrahedral arrangement. Only oxygen atoms are shown for clarity.
Each of these units corresponds effectively to a ramp particle. 
The hexagonal ice (the low pressure crystal of water) 
can be obtained by combining these units in an hcp lattice (the low pressure crystal of the ramp potential model).
We notice this figure is in the spirit of Walrafen pentamer~\cite{walrafen64} with the difference that the former
consists of only two water molecules.
}
\label{sergey}
\end{figure}

To define $m$ in physical units, we argue that the ramp particle
corresponds to two water molecules. This is based on the  
crystalline phases of water (hexagonal ice) and ramp potential (hcp) [see Fig.~\ref{pot}]. 
The hexagonal ice can be formed by combining units such as that shown in Fig.~\ref{sergey}. Each of this units is formed 
by a central water molecule and $1/4$ of its four neighbors, which are tetrahedrally arranged. To form the hexagonal ice, 
such units must form an hcp network. Therefore, if the crystalline structure of the ramp potential model is identified 
with that of hexagonal ice, a ramp particle must be identified, on average, to the unit shown in Fig.~\ref{sergey}. 
The mass of a water molecule is $m_w = 18$~g/mol, thus, the mass of a ramp particle 
is $m \approx (1 + 4 \times 1/4)~m_w= 36$~g/mol. 
Alternatively, the present argument implies that the number density of the ramp potential model corresponds to twice the number density of 
water, and this will be relevant when comparing the pressures of the ramp and TIP5P models~\cite{footnoteMassDouble}.
To test the idea that a ramp particle corresponds approximately to two water molecules, we calculate the average number of neighbors, $N_0$,
that a water molecule has within a distance of $r< \sigma_0=0.267$~nm. 
Using the $g(r)$ from Fig.~\ref{pot} we get $N_0 \equiv 4 \pi n \int_0^{\sigma_0} r'^2 g(r') dr' \approx 1$ (here, $n$ is the number density),
in agreement with our view. 
The correspondence between one ramp particle and two water molecules is also supported by
computer simulations of the ramp potential with $\lambda=0.581$
and an attractive part~\cite{wilding02,xu06}. Such a ramp potential model has 
both liquid-gas (LG) and liquid-liquid (LL) critical points. 
Application of the values for $\sigma_0$, $U_0$, and $m$ that we
use here to the data from~\cite{wilding02,xu06} results in $\rho_{LG}\approx0.314$ 
and $\rho_{LL}\approx1.188$~g/cm$^3$. 
These values approximately coincide with the experimental critical density of water
$\rho_{LG}\approx0.322$~\cite{critp1} and the LL critical density of TIP5P water $\rho_{LL}\approx1.13$~g/cm$^3$~\cite{critp2}. 

To compare the regions of anomalies in the phase diagrams of the TIP5P and ramp potentials, we
first obtain the LLCP coordinates,  $(P_{C},T_{C},\rho_{C})$. The LLCP in the TIP5P model 
is accessible in MD simulations and is  
located at $Tc=217$~K, $Pc=340$~MPa, and $\rho_c=1.13\pm0.04$~g/cm$^3$~\cite{yamada02,paschek}. 
Instead, for the ramp potential of Fig.~\ref{pot}(a), the LLCP is located at temperatures
below those accessible in simulations~\cite{predp05}.
In this case, the LLCP can be located by extrapolating the isochores in the $P-T$ phase diagram to low-$T$ (the isochores 
cross each other at the LLCP). 
This procedure indicates that the LLCP is located 
at $Tc=16.5$~K, $Pc=967$~MPa, and $\rho_c=1.19$~g/cm$^3$.

\begin{figure}
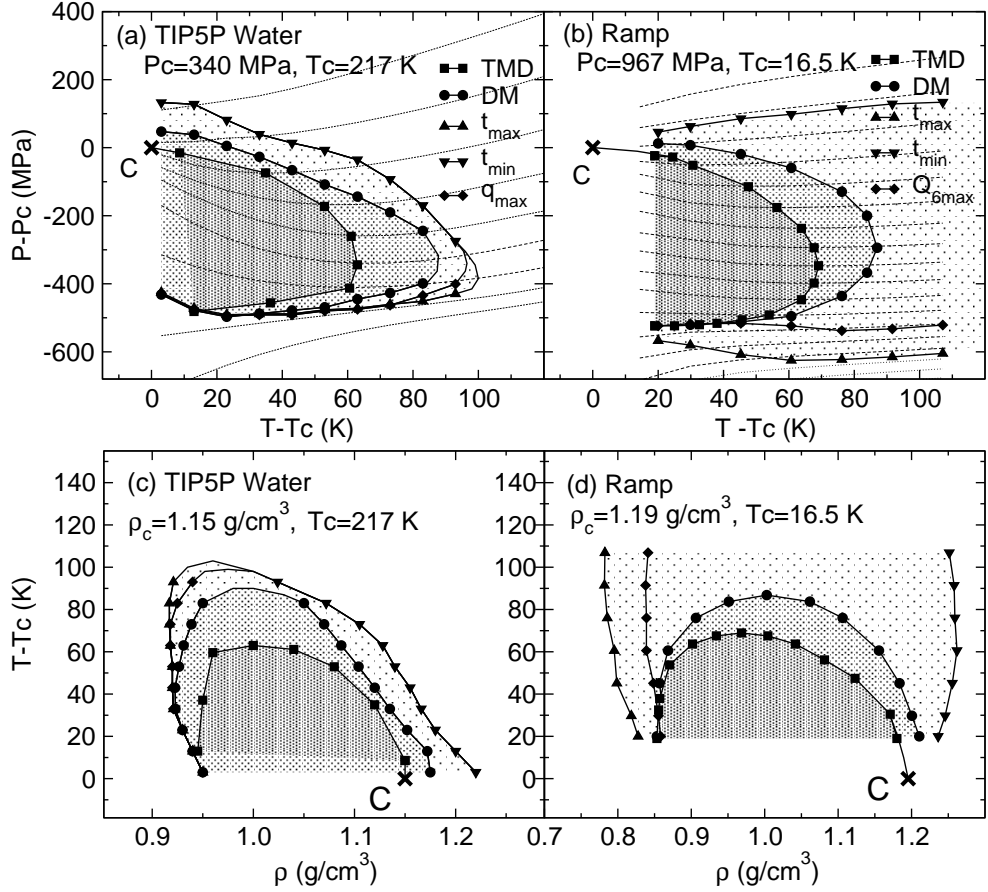

\includegraphics[width=13cm]{pttmddmqt7.eps}
\includegraphics[width=13cm]{tmddmqt9.eps}
\caption{
(a) Three anomalous regions of TIP5P water in modified $P$-$T$ phase diagram.
The dashed lines are the isochores with density $\rho$=1.20, 1.16, 1.12, 1.08, 1.04, 1.00,
0.96, 0.92, 0.88~g/cm$^3$ from top to bottom.
Density anomaly region is defined by TMD (temperature of maxima density) lines,
inside which the density increases when the system is heated at constant pressure.
Diffusion anomaly region is defined by the loci of DM (diffusion maxima-minima),
inside which the diffusivity increases with density at constant $T$.
Structural anomaly region is defined by the loci of translational order
minima ($t_{\min}$) and maxima ($t_{\max}$), or orientational order maxima $q_{\max}$ ($Q_{6max}$ for ramp liquid),
inside which both translational and orientational orders decrease with density at constant $T$ (see refs.~\cite{jeffrey01,zyan,zyan2007} 
for details).
Here $t$ quantifies the tendency of molecular pairs to adopt preferential separations, and
$q$ quantify the local tetrahedrality of water ($Q_6$ quantify the local orientational order of twelve nearest
neighbors in the first shell of a ramp particle).
(b) Anomalous regions for the ramp liquid, here the values of
$P$ and $\rho$ are doubled in order to compare with the corresponding values of water [see text].
The dashed lines are the isochores with density $\rho$=1.33, 1.28, 1.23, 1.18, 1.14,
1.09, 1.05, 1.02, 0.98, 0.94, 0.91, 0.88, 0.85, 0.82, 0.79, 0.77, 0.74
~g/cm$^3$ from top to bottom. The structural anomaly region is open and
converge at much higher $T$ compared with water.
(c) and (d) are the anomalous regions in the $T$-$\rho$ phase diagrams.}
\label{tmddmqt}
\end{figure}

Figure~\ref{tmddmqt} shows the phase diagrams
of the TIP5P and ramp potential models, obtained by MD simulations (for 
details see ~\cite{zyan2007,predp05, zyan}).
To emphasize the quantitative similarities of these diagrams we place the origins of $P$ and $T$ axes at
the LLCP of the corresponding models. 
In both models, the density anomaly region is within the diffusion anomaly region, which is
enclosed by the structure anomaly region.
A comparison of panels (a) and (b), or (c) and (d), shows {\it quantitative}
similarities in the regions of anomalies of both models. 
The density anomaly region covers approximately the ranges $-500<P-P_{C}<0$~MPa,
$T-T_{C}<60$~K, and $0.95< \rho <1.15$~g/cm$^3$, for the TIP5P model, and  $-520<P-P_{C}<-10$~MPa, $T-T_{C}<70$~K, and
$0.85< \rho <1.20$~g/cm$^3$ for the ramp potential model. 
Similarly, the diffusion anomaly region covers approximately the ranges $-500<P-P_{C}<50$~MPa, $T-T_{C}<90$~K, 
and $0.92< \rho<1.17$~g/cm$^3$ for the TIP5P model, and $-520<P-P_{C}<0$~MPa, $T-T_{C}<90$~K, and
$0.85< \rho<1.21$~g/cm$^3$ for the ramp potential model. The structure anomaly region, defined
by the loci of the order parameter extrema $(t_{min},q_{max})$ and $(t_{min},Q_{6max})$ for the TIP5P and ramp potential 
models respectively,
covers approximately the range $-500<P-P_{C}<150$~MPa and $0.9<\rho<1.22$~g/cm$^3$, for the TIP5P model, 
and $-500<P-P_{C}<150$~MPa and $0.85<\rho<1.27$~g/cm$^3$, for the ramp potential model.
However, the shape of the structural anomaly
region is different in these models. It expands up to $T-T_{C}<100$~K for the TIP5P model, while 
for the ramp potential it expands to much higher temperature outside the graph.

\begin{figure}
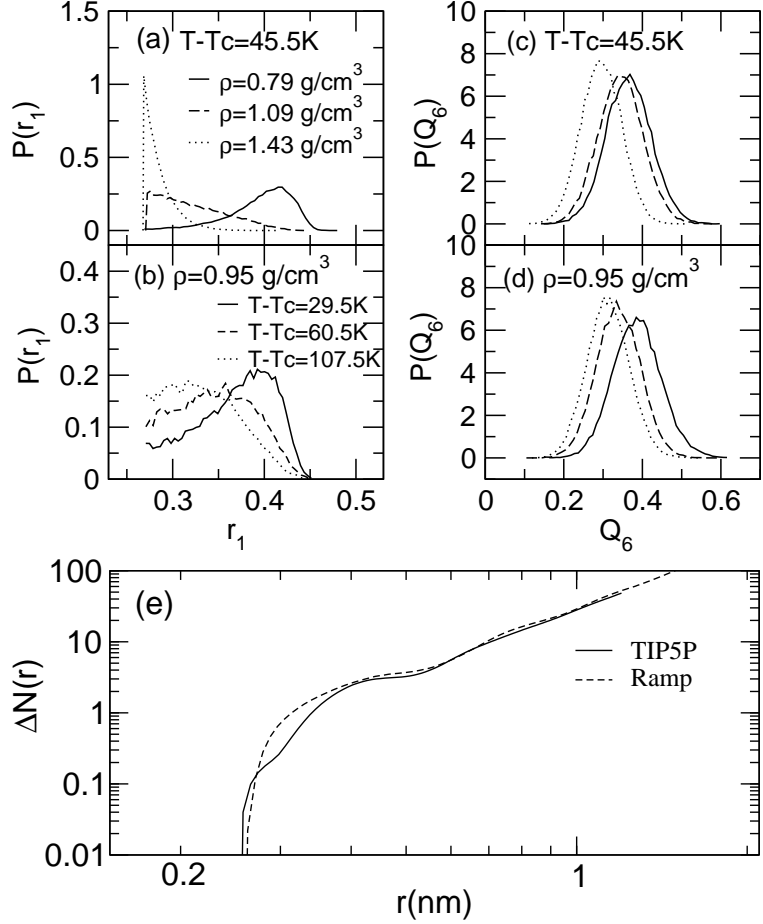

\includegraphics[width=10cm]{his6.eps}
\includegraphics[width=10cm]{dn.eps}
\caption
{Probability distribution of the distance $r_1$ of a central ramp particle and its nearest neighbor at (a)
constant $T$ and (b) constant $\rho$. (c)-(d) Probability distribution of orientational order parameter for a ramp 
potential particle corresponding to panels (a) and (b), respectively.
Upon heating or compression, ramp particles move from the soft-core distance toward the hard-core distance and the orientation order parameter 
decreases. Similar structural changes occur in water~\cite{zyan2007}.
(e) Increase in the number of neighbors,  $\Delta N(r) \equiv N(r)|_{\rho_1} -  N(r)|_{\rho_0}$, where $\rho_1=0.88$~g/cm$^3$ and $\rho_0=1.08$~g/cm$^3$,
for the TIP5P and ramp potentials. 
$\Delta N(r)$ in both models are remarkably similar and overlap for approximately 
$r >0.37$~nm. Since a ramp particle corresponds to two water molecules,
we doubled the values of $N(r)$ and $\Delta N(r)$ obtained from the simulations using the ramp potential model.
}
\label{his}
\end{figure}

A possible reason for the {\it quantitative} similarities in the regions of anomalies of water and ramp potential model 
is that this model is able to reproduce {\it quantitatively} the observed migration of water molecules
from the second shell toward the first shell upon compression or heating~\cite{ssitta03,zyan2007,scio90}.
We discuss first the probability distribution, $P(r_{1})$, of the distance between a ramp particle
and its nearest-neighbor. Figure~\ref{his}(a) shows the evolution of $P(r_{1})$ upon isothermal 
compression.  As density increases, the maxima of $P(r_{1})$ shifts from $r=0.42 \approx \sigma_1$, at low density, to
$r=0.267 = \sigma_0$, at high density. Figure~\ref{his}(b) shows that a similar but less pronounced changes in $P(r_{1})$ occur
upon isobaric heating. Thus, upon compression or heating, particles move from the soft-core distance (corresponding to
water's second shell) toward the hard-core distance (corresponding to water's first shell) of the ramp potential. 
Similar structural changes occur in water~\cite{ssitta03,zyan2007,scio90}. In particular, Figures~\ref{his}(a)
and~\ref{his}(b) can be compared with the corresponding Figs.~4(c) and 4(f) of ref.~\cite{zyan2007} obtained for the TIP5P model. 
The probability distribution, $P(Q_{6})$, of the orientational order parameter, $Q_{6}$~\cite{zyan}, of the ramp potential
particles is shown in Figs.~\ref{his}(c) (upon isothermal compression) and~\ref{his}(d) (upon isobaric heating).  
Upon compression or heating, the maximum  $P(Q_{6})$ shifts to small values of $Q_{6}$, i.e. orientational order decreases.  
Similar structural changes occur in the orientational order of water's second shell; Figs.~\ref{his}(c)
and~\ref{his}(d) can be compared with the corresponding Figs.~4(b) and 4(e) of ref.~\cite{zyan2007} obtained for the TIP5P model. 
For a quantitative comparison of the structural changes in the ramp and TIP5P models, 
we calculate the number of neighbors, $N(r)$, as a function of the distance $r$ from a central water molecule/ramp particle in both models.
The increase of $N(r)$ with density, $\Delta N(r)$, is shown in Fig.~\ref{his}(e).
We see that $\Delta N(r)$ for both models are remarkably similar and overlap for approximately 
$r >0.37$~nm. Thus, the ramp potential reproduces  {\it quantitatively} the migration of water molecules
from the second shell toward the first shell upon compression or heating.

In summary, our study makes a microscopic {\it quantitative} connection between
a ramp potential and TIP5P water model and shows that 
orientational interactions, such as hydrogen bonding, are not necessary to 
reproduce water-like anomalous properties. In general, the ramp potential provides an
understanding of the anomalous features of tetrahedral liquids. These features are caused by
a large empty space around the tetrahedrally coordinated molecules, which is reduced as
temperature and pressure increase. In the ramp liquid, this empty space is created by the repulsive soft core.

We thank NSF for support with grants CHE~0096892. SVB gratefully acknowledges the partial support of this research through the Dr.
Bernard W. Gamson Computational Laboratory at Yeshiva College.
We are thankful to P. G. Debenedetti for enlightening discussion and for comments on
the manuscript.

\clearpage

\newpage

\end{document}